\documentstyle[aps,prl,epsf,floats,cite]{revtex}

\def\lsim{\mathrel{\raise.3ex\hbox{$<$\kern-.75em\lower1ex\hbox{$\sim$}}}}
\def\gsim{\mathrel{\raise.3ex\hbox{$>$\kern-.75em\lower1ex\hbox{$\sim$}}}}

\newcommand{\nn}{\nonumber}

\newcommand{\bmp}{\mbox{\boldmath $p$}}
\newcommand{\bmq}{\mbox{\boldmath $q$}}
\newcommand{\bmk}{\mbox{\boldmath $k$}}

\def\O#1#2{\mbox{\boldmath $O$}_{\mbox{\scriptsize\boldmath $#1$},#2}}

\def\bsigma{\mbox{\boldmath $\sigma$}}
\def\lqcd{\Lambda_{\rm QCD}}

\def\psip#1{\psi_{\mbox{\scriptsize\boldmath $#1$}}}
\def\chip#1{\chi_{\mbox{\scriptsize\boldmath $#1$}}}

\def\OMIT#1{}

\begin{document}

\twocolumn[\hsize\textwidth\columnwidth\hsize\csname@twocolumnfalse\endcsname

\mbox{}\hfill
\vbox{\tighten 
\hbox{CERN-TH/2000-340} \hbox{UCSD/PTH 2000-25} 
                 \hbox{PITHA 00/28}
                 \hbox{hep-ph/0011254} 
                 \hbox{} } 

\title{A Renormalization Group Improved Calculation of \\
Top Quark Production near Threshold  } 

\author{ A.~H.~Hoang${}^a$, A.~V.~Manohar${}^b$,
         I.~W.~Stewart${}^b$ and T.~Teubner${}^c$\\[3mm]}

\address{${}^a$ Theory Division, CERN,
   CH-1211 Geneva 23, Switzerland}
\address{${}^b$ Department of Physics, UCSD,
  9500 Gilman Drive, La Jolla, CA 92093-0319, USA}
\address{${}^c$ Institut f\"ur Theoretische Physik E, RWTH Aachen,
   D-52056 Aachen, Germany}
\maketitle

{\tighten
\begin{abstract}
The top quark cross section close to threshold in $e^+e^-$ annihilation is
computed including the summation of logarithms of the velocity at
next-to-next-to-leading-logarithmic order in QCD.  The remaining theoretical
uncertainty in the normalization of the total cross section is at the few
percent level, an order of magnitude smaller than in previous
next-to-next-to-leading order calculations.  This uncertainty is smaller than
the effects of a light standard model Higgs boson.
\end{abstract}
\pacs{}
}
]\narrowtext 

Top quark pair production close to threshold is a major part of the top quark
physics program at a future lepton collider. Near threshold the top quark
velocity $v$ is small and the presence of Coulomb singularities make the
summation of terms proportional to $\alpha_s/v$ mandatory.  The top quark width,
$\Gamma_t\approx 1.5\,\mbox{GeV}\gg\lqcd$, serves as an infrared cutoff,
allowing for the use of perturbative methods to calculate the non-relativistic
top-antitop dynamics to a high degree of precision.

Recent NNLO QCD calculations of the total cross section $\sigma_{t\bar t}$
showed that the top quark mass can be determined with uncertainties below
$200$~MeV, but indicated that the strong coupling, the top Yukawa coupling, and
$\Gamma_t$ cannot be measured with good precision due to large theoretical
normalization uncertainties of about $20\%$~\cite{Hoang3}.  A common feature of
all recent NNLO QCD calculations is that they are fixed order calculations,
i.e.\ the running from the hard scale $m_t$ down to the non-relativistic scales
which govern the dynamics of the top-antitop system was not taken into
account. For $t\bar t$, $v\sim 0.15$ so that logarithms of ratios of $m_t \sim
175$~GeV, $m_t v \sim 25$~GeV and $m_t v^2 \sim 4$~GeV are not small, and the
renormalization group evolution is significant.

In this letter we calculate the photon induced $t\bar t$ production
cross-section in the framework of vNRQCD using the velocity renormalization
group (VRG)~\cite{Luke1}. The calculation includes a summation of logarithms of
the ratios of the scales $m_t$, $m_t v$ and $m_t v^2$ at next-to-next-to-leading
order. The VRG-improved computation has a scale uncertainty of 2--3\% at the
peak of the cross section (without initial state radiation and beam smearing
effects). An improvement in the convergence of the expansion is also found.
Measurements of the strong coupling, the top Yukawa coupling, and the top quark
width appear feasible with small theoretical uncertainties.

The expansion for the normalized cross section  $R=\sigma_{t\bar
t}/\sigma_{\mu^+\mu^-}$ takes the form
\begin{eqnarray}
 R =
 v\,\sum\limits_{k,i} \left(\frac{\alpha_s}{v}\right)^k
 \left(\alpha_s\ln v \right)^i \left\{
 \begin{array}{ll}
 1\, & \mbox{(LL)} \, \\
 \alpha_s, v\, & \mbox{(NLL)} \\
 \alpha_s^2, \alpha_s v, v^2\, & \mbox{(NNLL)}
 \end{array} \right. .
 \label{RNNLLorders}
\end{eqnarray}
The free quark cross-section is of order $v$. The Coulomb summation of powers of
$\alpha_s/v$, and the VRG summation of powers of $\alpha_s \ln v$, are the sums
over $k$ and $i$, respectively. Terms in the cross section at 
leading-logarithmic (LL), next-to-leading-logarithmic (NLL), and 
next-to-next-to-leading-logarithmic (NNLL) order are indicated in 
Eq.~(\ref{RNNLLorders}).

vNRQCD~\cite{Luke1} is an effective field theory which describes
non-relativistic heavy quarks with mass $m$ interacting with soft gluons with
four-momenta $k^\mu\sim m v$ and ultrasoft gluons with $k^\mu\sim m
v^2$, where $m v^2$ is larger than $\lqcd$. For
loop integrations over soft energies and momenta the 
$\overline{\mbox{MS}}$ subtraction scale is $\mu_S=m \nu$, whereas for
loop integrations with ultrasoft energies and momenta it is $\mu_U=m
\nu^2$. The  
subtraction velocity $\nu$ correlates $\mu_S$ and $\mu_U$ and is used instead of
the $\overline{\mbox{MS}}$ scale parameter $\mu$.  The correlation is mandatory
since energy and momentum are related through the quark equations of motion.  In
QED this has been shown to be necessary to reproduce through running the
$(\ln\alpha)^k$, $k\ge 2$ contributions in Lamb shifts, hyperfine splittings and
corrections to the ortho and para positronium decay widths~\cite{Manohar5}.

Lowering $\nu$ from $\nu=1$ to the quark velocity $v$, sums all large logarithms
involving the soft and ultrasoft scales into the Wilson coefficients of the
vNRQCD operators. At NNLL this includes logarithms originating from radiation
effects.  Once $\nu$ is lowered to $v$, power counting shows that matrix
elements with ultrasoft gluons do not have to taken into account at NNLL for the
description of a heavy quark pair in a color singlet state.  Thus, after
lowering $\nu$ to order $\alpha_s$ (since $v\sim \alpha_s$ in a Coulombic
system) the NNLL equation of motion of a color singlet quark-antiquark system
is a conventional two-body Schr\"odinger equation. 

In momentum space the NNLL Schr\"odinger equation reads
\begin{eqnarray} 
 && 
 \bigg[ \frac{\bmp^2}{m} - \frac{\bmp^4}{4m^3} - E \bigg]
 \, \tilde G(\bmp,\bmp^\prime) + \int\!D^n\bmq\, 
 \mu_S^{2\epsilon}\tilde V(\bmp,\bmq)\, 
 \tilde G(\bmq,\bmp^\prime) \nn \\
 &&
\hspace{1cm}
= \, (2\pi)^n\,\delta^{(n)}(\bmp-\bmp^\prime) \,,
\label{NNLLSchroedinger}
\end{eqnarray}
where $m$ is the heavy quark pole mass, $E\equiv (\sqrt{s}-2m)$, $n\equiv
3-2\epsilon$, and $D^n\mbox{\boldmath $q$}\equiv e^{\epsilon\gamma_E}
(4\pi)^{-\epsilon}\,d^n\bmq/(2\pi)^n$. 
The potential for the quark-antiquark pair in a ${}^3S_1$ state (relevant
for top production through a virtual photon) is
\begin{eqnarray}
 \tilde V(\bmp,\bmq) \, = \, 
 \tilde V_c(\bmp,\bmq) +
 \tilde V_k(\bmp,\bmq) + 
 \tilde V_\delta(\bmp,\bmq) + 
 \tilde V_r(\bmp,\bmq)  
 \,,
\label{Vsdetail}
\end{eqnarray} 
where 
($a_s=\alpha_s(\mu_S)$, $L=\ln(\bmk^2/\mu_S^2)$, 
$\bmk=\bmp-\bmq$)
\begin{eqnarray}
 \tilde V_c(\bmp,\bmq) 
 & = &
 \frac{{\cal{V}}_c(\nu)}{\bmk^2}
 -\frac{4\pi C_F a_s}{\bmk^2}\bigg\{
 \frac{a_s}{4\pi}\,\Big[-\beta_0 L + a_1\Big] \nonumber\\
& &+ \big(\frac{a_s}{4\pi}\big)^2\Big[
 \beta_0^2 L^2 - (2 \beta_0 a_1 + \beta_1) L + a_2 \Big]\bigg\} \,,
 \label{VCoulomb}
\\
 \tilde V_k(\bmp,\bmq)
 & = &
 \frac{\pi^2}{m |\bmk|}\, {\cal{V}}_k(\nu) \,,
 \label{Vk}
\\
 \tilde V_\delta(\bmp,\bmq)
 & = & \frac{{\cal{V}}_2(\nu) +2 {\cal{V}}_s(\nu)}{m^2}\,,
 \label{Vdelta}
\\
 \tilde V_r(\bmp,\bmq)
 & = &
 \frac{(\bmp^2+\bmq^2)}{2 m^2 \bmk^2}\, {\cal{V}}_r(\nu) \,.
 \label{Vr}
\end{eqnarray}
The potentials arise from the four-quark matrix elements of potential-type
operators and from time-ordered products of operators describing interactions
with soft gluons. The coefficients of the potentials at the hard scale, ${\cal
V}_i(\nu=1)$, are obtained with on-shell matching. For $\nu\sim v$ the
coefficients contain the summation of all $\ln v$ terms.  The velocity counting
of each potential in Eq.~(\ref{NNLLSchroedinger}) is equivalent to the counting
in the Schr\"odinger equation in previous NNLO calculations (see e.g.\ Ref.\
\cite{Hoang2}). The coefficient ${\cal V}_c(\nu)$ in the Coulomb potential,
$\tilde V_c$, was determined in Ref.\ \cite{Manohar1} at NNLL order. 
The second
term in Eq.~(\ref{VCoulomb}) contains the one and two-loop corrections to the
Coulomb potential~\cite{Schroder1}. In vNRQCD it arises from the time-ordered
product of the lowest order operators describing the interaction of quarks with
soft gluons. The couplings in this time ordered product are 
$\alpha_s(\mu_S)$ and evolve with the QCD $\beta$-function.  The potential
$\tilde V_k$ leads to terms in the cross section that are $v^2$-suppressed. The
coefficient ${\cal V}_k$ has a matching value of order
$\alpha_s^2$~\cite{Titard1}, so its NLL evolution from Ref.\ \cite{Manohar3} is
needed. The potentials $\tilde V_\delta$ and $\tilde V_r$ also lead to terms in
the cross section that are $v^2$-suppressed and have order $\alpha_s$
coefficients ${\cal V}_{2,s,r}$ generated at tree level. Their evolution is only
needed at LL order and can be found in Ref.\ \cite{Manohar2}.

To describe vector current induced quark-antiquark production close to threshold
at NNLL order we also need the Wilson coefficients of the dimension-three
${}^3S_1$ currents at NNLL order and the corresponding dimension-five currents
at LL order. The non-relativistic current for production is ${\bf J}_{\bf p}=
c_1 \O{p}{1} + c_2 \O{p}{2}$, where
\begin{eqnarray}
 \O{p}{1} & = & {\psip{p}}^\dagger\,\bsigma\,(i\sigma_2)\,{\chip{-p}^*} \,, \\
 \O{p}{2} & = & \frac{1}{m^2}\, {\psip{p}}^\dagger\, 
  \big( \bmp^2 \,\bsigma \big)\, (i\sigma_2)\,{\chip{-p}^*} \,.
\end{eqnarray} 
Spin and color indices are suppressed. There is another dimension-five current
describing D-wave production which does not contribute at this order.  The
corresponding annihilation currents ${\O{p}{1,2}}^\dagger$ are obtained by
complex conjugation. The matching condition for the Wilson coefficient $c_1$ at
the hard scale needs to be known at order $\alpha_s^2$, and the Wilson
coefficient $c_2$ needs to be known at the Born level.  The value of
$c_1(\nu=1)$ can be determined from matching the two-loop result for the
quark-antiquark production amplitude close to threshold in full
QCD~\cite{Czarnecki1} to the corresponding amplitude in vNRQCD. We find
\begin{eqnarray}
c_1(1) & = & 
 1-2 C_F \frac{\alpha_s(m)}{\pi} +   
\alpha_s^2(m) \bigg[C_F^2\Big(\frac{\ln 2}{12}-\frac{25}{24}
 -\frac{2}{\pi^2}\Big) \nonumber\\ & & 
+ C_A C_F\Big(\ln 2-1\Big) + \frac{\kappa}{2} \bigg] \,,
\\
\kappa & = & C_F^2 \bigg[ \frac{1}{\pi^2}\Big(\frac{39}{4}-\zeta_3\Big) +
 \frac{4}{3}\ln 2 - \frac{35}{18} \bigg] \nonumber\\ 
 & & - C_A C_F \bigg[ \frac{1}{\pi^2} \Big(
 \frac{151}{36} + \frac{13}{2} \zeta_3 \Big) +
 \frac{8}{3} \ln 2 - \frac{179}{72} \bigg] \nonumber\\ 
 & & + C_F T_F \bigg[ \frac{4}{9} \Big( \frac{11}{\pi^2} - 1 \Big) \bigg] +
C_F T_F n_l \bigg[ \frac{11}{9 \pi^2} \bigg] 
\,.
\end{eqnarray}
The two-loop result for $c_1(1)$ is scheme-dependent and our result differs from
the hard contribution obtained from the threshold expansion\
\cite{Czarnecki1,Beneke2}. 

The LL anomalous dimension for $c_1$ is zero.  The evolution of $c_1$ for
$\nu<1$ at NLL order has been determined analytically in Ref.\ \cite{Manohar3}
by solving~\cite{Luke1}
\begin{eqnarray} \label{adc1}
 \frac{\nu}{c_1} \frac{\partial}{\partial\nu} c_1
 &=& -\frac{{\cal V}_c}{16\pi^2} 
\Big( \frac{{\cal V}_c}{4}
  +{\cal V}_2+{\cal V}_r
  + 2 {\cal V}_s  \Big) +
  \frac{{\cal V}_{k}}{2} \,. 
\end{eqnarray}
The resummed logarithms in $c_1(\nu=\alpha_s)$ at NLL order include the sizeable
negative normalization corrections $\propto\alpha_s^3\ln^2\alpha_s$ found in
Ref.\ \cite{Kniehl1}. At NNLL order we find that there are no additional
contributions to the anomalous dimension for $c_1$, so Eq.~(\ref{adc1}) remains
valid.  However, solving for the full NNLL $c_1(\nu)$ requires the NLL values of
${\cal V}_c$, ${\cal V}_2$, ${\cal V}_r$ and ${\cal V}_s$, and the NNLL value of
${\cal V}_k$. Besides ${\cal V}_c$, the coefficients at this order are
not yet known. For our NNLL solution we will use $c_1(1)$ at ${\cal
O}(\alpha_s^2)$  with NLL evolution in
$\nu$.  From previous experience in weak decays, it is not expected that the NLL
(NNLL) results for ${\cal V}_2$, ${\cal V}_r$ and ${\cal V}_s$ (${\cal V}_k$)
will deviate significantly from the LL (NLL) ones.  Thus, our result for the
top-antitop cross section should yield a realistic estimate of the theoretical
uncertainties at NNLL order.  The LL result for $c_2$ is generated by ultrasoft
gluons renormalizing $\O{p}{1}$ and reads
\begin{eqnarray}
 c_2(\nu)&=& -\frac{1}{6}-
 \frac{32}{\beta_0}\ln\Big(\frac{\alpha_s(m\nu^2)}{\alpha_s(m)}\Big)
 \,.
\end{eqnarray}
The first logarithm in this series agrees with the logarithm in the matching
calculation in Ref.~\cite{Luke2}.

In full QCD the expression for the normalized photon-induced cross
section for heavy top-antitop production $R=\sigma_{t\bar
t}/\sigma_{\mu^+\mu^-}$ at the c.m.\ energy $\sqrt{s}$ is
\begin{equation}
 R \, = \,
 \frac{16 \pi}{9 s}\,\mbox{Im}\,\left[-i
 \int\! d^4x \,e^{i q.x} \left\langle 0 \left|T j^\mu(x) 
 j_\mu(0) \right| 0 \right\rangle \right] \,,
\end{equation}
where $q=(\sqrt{s},0)$ and $j^\mu$ is the vector current that produces a
top-antitop pair. In vNRQCD at NNLL order the vector current correlator is
replaced by the correlators of the non-relativistic currents $\O{p}{1}$ and
$\O{p}{2}$ evaluated for $\nu\sim\alpha_s$. The correlator of two $\O{p}{1}$
currents is proportional to the coordinate space Green function $G(0,0)$
obtained from the NNLL Schr\"odinger equation~(\ref{NNLLSchroedinger}) and the
correlator of $\O{p}{1}$ and $\O{p}{2}$ is proportional to $(E/m)G(0,0)$. To
determine $G(0,0)$ at NNLL order we use a combination of numerical and analytic
calculations~\cite{Hoang1}. The ultraviolet-finite contributions from the
Coulomb potential\ (\ref{VCoulomb}) (denoted by $G^c$ in Eq.\
(\ref{NNLLcrosssection}) below) are determined exactly using numerical
techniques developed in Ref.~\cite{Jezabek1}. The ultraviolet-divergent
contributions from $\tilde V_\delta$, $\tilde V_r$, $\tilde V_k$ and the kinetic
energy correction (denoted by $\delta G^\delta$, $\delta G^r$, $\delta G^k$ and
$\delta G^{\rm kin}$, respectively) are computed in perturbation theory with
dimensional regularization in the $\overline{\rm MS}$ scheme. This is required
for consistency with the scheme used to compute the matching and running of the
Wilson coefficients. By power counting, only a single insertion of these
potentials is included.  In deriving these results we have included the
counter terms generated by renormalizing the $\O{p}{1}$
current~\cite{Luke1}. These counter term graphs are sufficient to cancel all
subdivergences.  The remaining overall divergences are of the form $1/\epsilon$
and $v^2/\epsilon$ and are cancelled by renormalizing the time ordered product
of currents.  The final result for the NNLL cross section in vNRQCD is
\begin{eqnarray}
\lefteqn{
 R^{\rm NNLL}
 \, = \,
 \frac{8 \pi}{m^2}\, \mbox{Im}\Big\{\,  
 c_1^2\,\Big[\,(1-v^2)\,G^c
 + ({\cal{V}}_2+2{\cal{V}}_s)\,\delta G^\delta
}
\nonumber
\\ & & \hspace{6mm}
+ {\cal{V}}_r\,\delta G^r
+ {\cal{V}}_k\, \delta G^k 
+ \delta G^{\rm kin} 
\,\Big]
 + 2 c_1\,c_2\,v^2\,G^c
\Big\}
\,,
\label{NNLLcrosssection}
\end{eqnarray}
where $v$ is the top quark velocity and the dependence of the Wilson
coefficients on $\nu$ is suppressed.  The top quark width is implemented by
shifting the energy into the positive complex energy plane by
$i\Gamma_t$~\cite{Fadin1}, which leads to the following expression for the top
quark velocity
\begin{eqnarray}
v & = &
\left(\frac{\sqrt{s}-2m_t+i\Gamma_t}{m_t}\right)^{\frac{1}{2}} \,.
\label{vdefwidth}
\end{eqnarray} 
We emphasize that Eq.\ (\ref{vdefwidth}) does not
provide a consistent treatment of the top quark width beyond next-to-leading
order.  This can be seen from the presence of $\nu$-dependent terms proportional
to $\alpha_s\Gamma_t/m_t \ln(-iv/\nu)$ in Eq.\ (\ref{NNLLcrosssection}) which
are parametrically of NNLL order\ \cite{Hoang1}.  Conceptually, they indicate
that further renormalization procedures are required in a consistent treatment
of electroweak effects.
 
%
%

%
%
\begin{figure}[t] 
\begin{center}
 \leavevmode
 \epsfxsize=3.5cm
 \epsffile[220 580 420 710]{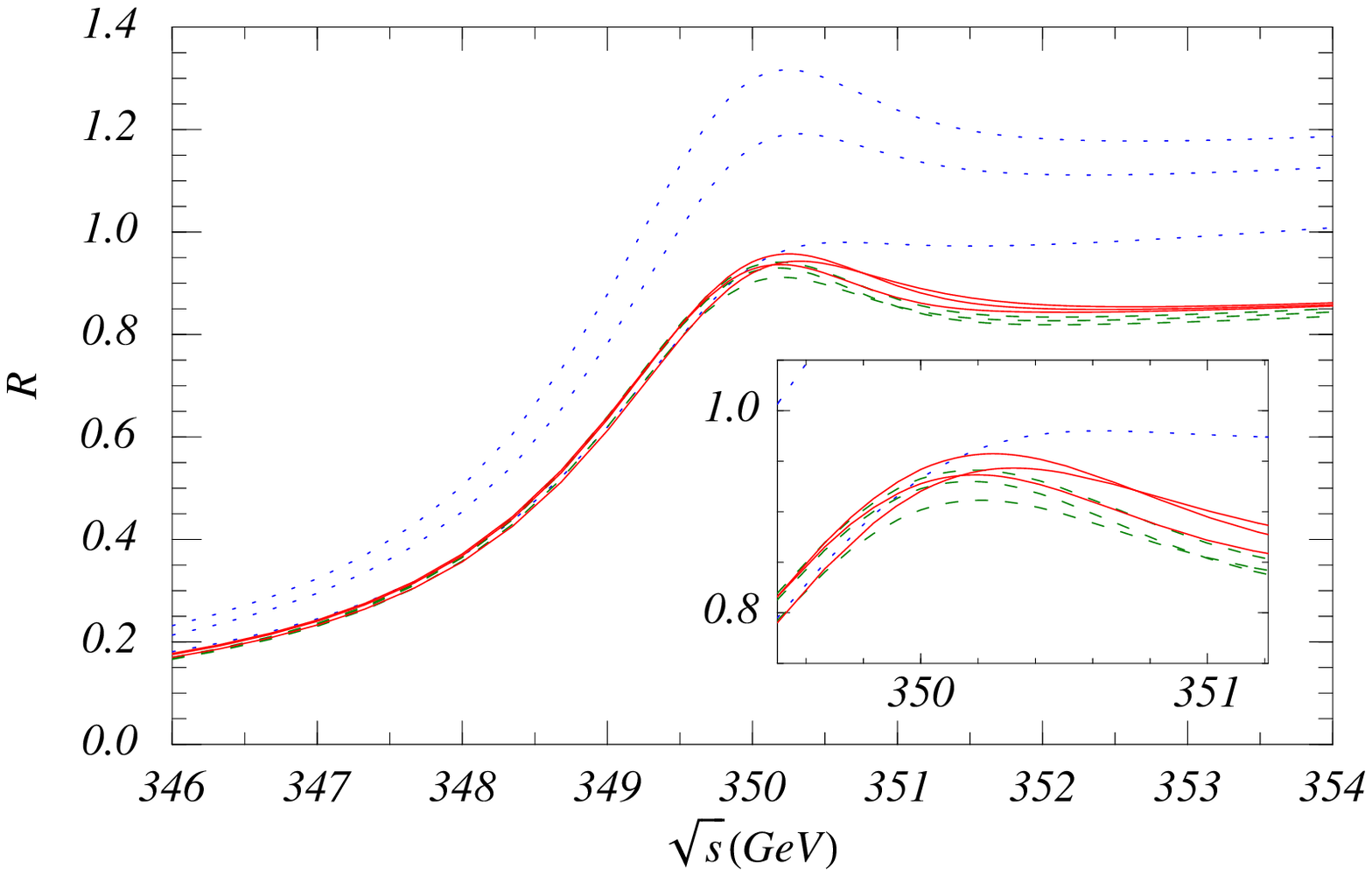}
\\[3.1cm]
 \leavevmode
 \epsfxsize=3.4cm
 \epsffile[220 580 420 710]{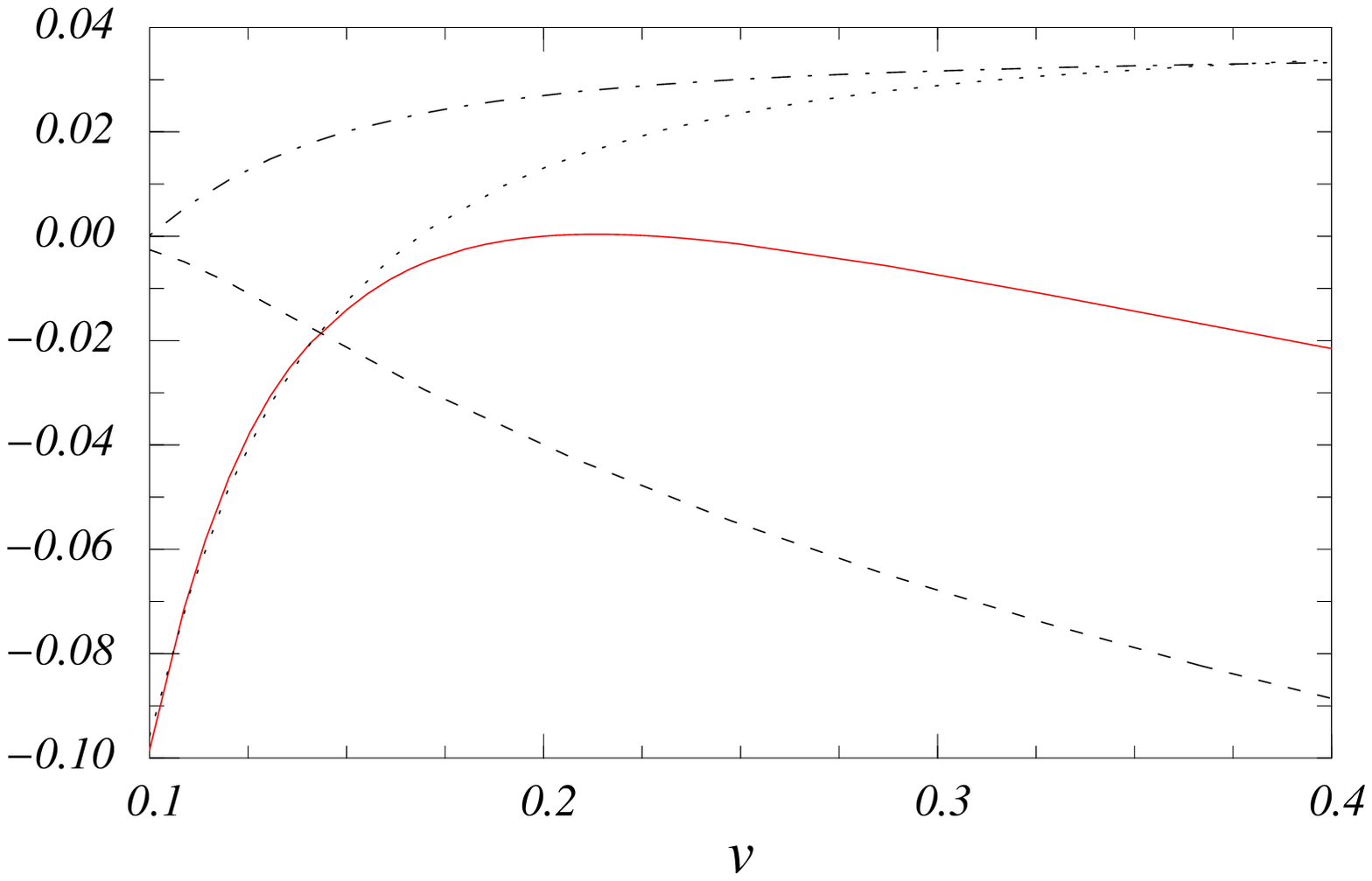}
\vskip 2.7cm 
\caption{ The upper panel shows the normalized photon-induced cross
section for $t\bar t$ at LL order (dotted), NLL order (dashed), and NNLL order
(solid) in the 1S mass scheme with $\nu=0.15$, $0.2$, and $0.4$. 
The lower panel shows the $\nu$ dependence
of the contributions in Eq.\ (\ref{NNLLcrosssection}) 
to $R^{\rm NNLL}(\sqrt{s}=350\,{\rm GeV})$. The $\delta G^k$
contribution is shown by the dotted line, 
the sum of $\delta G^{\delta,r,{\rm kin}}$ terms by the dot-dashed line, the
terms involving $\delta G^c$ by the dashed line, and the sum by the solid
line. \label{fig:nnllplots} } \end{center}
\end{figure}
In the upper panel of Fig.~\ref{fig:nnllplots} the normalized top quark cross
section is displayed in the 1S mass scheme~\cite{Hoang4,Hoang2} versus the
c.m.\ energy $\sqrt{s}$, for $m_t^{\mbox{\tiny 1S}}=175$~GeV,
$\alpha_s^{(n_f=5)}(M_Z)=0.118$ and $\Gamma_t=1.43$~GeV. The LL (dotted lines),
NLL (dashed lines), and NNLL (solid lines) results are shown for $\nu=0.15$,
$0.2$, and $0.4$.  From a physical point of view the appropriate choice of the
subtraction parameter $\nu$ is around $\alpha_s\approx 0.15$--$0.2$.  The LL
curves for $\nu=0.15, 0.2$ and $0.4$ correspond to the upper, middle and lower
lines, respectively. The NLL and NNLL curves differ so little on the vertical
scale of the figure that we refrain from labelling them.  At NNLL order, the
relative variation of the cross section at the peak position is 2\% for
$0.15<\nu<0.4$, whereas for $\sqrt{s}=(346,350,351,354)$~GeV the variation is
$(4\%,2\%,3\%,0.7\%)$.  At NLL and LL order the corresponding variation are
$3\%$, $(2\%,3\%,2\%,2\%)$ and $34\%$, $(28\%,39\%,27\%,18\%)$, respectively.
Overall, the variation of the normalization of the cross section for reasonable
choices of $\nu$, and the shift due to the NNLL order corrections are an order
of magnitude smaller than in previous NNLO calculations where threshold masses
were employed (see Ref.\ \cite{Hoang3}). The improved stability of the NNLL
cross section is a consequence of the evolution of the Wilson coefficients for
the potentials and the currents.

In the lower panel of Fig.~\ref{fig:nnllplots} the contributions to
the cross section at 
$\sqrt{s}=350\,{\rm Gev}$ coming from $G^c$ (dashed line), $\delta G^k$ (dotted
line), and the sum of $\delta G^{\delta,r,{\rm kin}}$ (dot-dashed line) are
displayed as a function of $\nu$.  The sum of all contributions is represented
by the solid line. For convenience of presentation we have subtracted the value
of $R^{\rm NNLL}$ at $\nu=0.2$ from the dashed and solid curves.
Whereas the individual 
contributions vary quite rapidly, most notably the Coulomb term $G^c$ and the
$1/|{\bf k}|$ potential term $\delta G^k$, there is a partial
cancellation in the sum, which varies more slowly. The correlation of
the Wilson coefficients that leads
to this stability is a consequence of the vNRQCD renormalization group equations
that account for the correlated soft and ultrasoft running and the mixing of
Wilson coefficients for $\nu<1$.  We note that the corrections coming from
$\tilde V_k$ quickly become negative for small $\nu$, and lead to an
instability of the NNLL curve at the peak if $\nu$ is chosen smaller than the
Coulombic velocity $v\simeq 0.15$. For $\nu>0.15$ multiple energy poles caused
by the perturbative treatment of the potentials $\tilde V_\delta$, $\tilde V_r$
and $\tilde V_k$ and the kinetic energy correction do not have to be resummed
because the corresponding corrections in the binding energies are an order of
magnitude smaller than the top quark width, $\Gamma_t\approx 1.5$~GeV. However,
for $\nu<0.15$ the residue of the double-pole ${\cal V}_k$ term becomes large and
multiple insertions of $\tilde V_k$ at the peak have to be summed to stabilize
the cross section. For $\nu<0.1$ the value of the Wilson coefficients changes
rapidly due to the fact that $\mu_U=m_t\nu^2$ gets close to $1$~GeV.

It is instructive to consider the size of some current correlator corrections
from beyond NNLL for $\nu>0.15$. The corrections arising from two insertions of
$\tilde V_\delta$ are smaller than $1$\% for $\nu>0.15$. In Ref.\ \cite{Kniehl2}
the corrections to the square of the heavy quarkonium (nS) wave function at the
origin arising from the emission and reabsorption of an ultrasoft gluon were
determined, which can be taken as an estimate for ultrasoft corrections to the
cross section. To be compatible with our calculations the $\overline{\mbox{MS}}$
parameter used in Ref.\ \cite{Kniehl2} has to be replaced by the ultrasoft scale
$\mu_U$. For the ground state ($n=1$) one finds that the corrections amount to
about $2$\% for $\nu\approx \alpha_s$. The small size of the two corrections
just mentioned strengthens our confidence that the 2--3\% variation of the
normalization with $\nu$ at NNLL order represents a realistic estimate of the
theoretical uncertainties.  Using the prescription given in
Ref.~\cite{Harlander2} the relative corrections to the normalization of the
cross section from a $115\,{\rm GeV}$ standard model Higgs boson are 5--8\% for
energies near the threshold. This is larger than the remaining uncertainty of
the NNLL cross section.

AH is supported in part by the EU Fourth Framework Program ``Training and
Mobility of Researchers'', Network ``Quantum Chromodynamics and Deep Structure
of Elementary Particles'', contract FMRX-CT98-0194 (DG12-MIHT). IS is supported
in part by NSERC of Canada, and AM and IS are supported in part by the
U.S.~Department of Energy under contract~DOE-FG03-97ER40546.

\end{document}